\documentclass[twocolumn,showpacs,aps,prl,superscriptaddress]{revtex4}

\usepackage{graphicx}
\usepackage{dcolumn}
\usepackage{amsmath}
\usepackage{epsfig}

\input pubboard/babarsym

\newcommand{\etal}{{\it et al.}}
\def\Btopipi   {\ensuremath{\B \to \pi\pi}\xspace}
\def\Btopipiz   {\ensuremath{\Bpm \to \pipm\piz}\xspace}

\def\Btokpiz   {\ensuremath{\Bpm \to \Kpm\piz}\xspace}
\def\Btohpiz   {\ensuremath{\Bpm \to \hpm\piz}\xspace}
\def\btokpiz   {\ensuremath{\Bpm \to \Kpm\piz}\xspace}
\def\btohpiz   {\ensuremath{\Bpm \to \hpm\piz}\xspace}
\def\btopipiz   {\ensuremath{\Bpm \to \pipm\piz}\xspace}
\def\Bztopizpiz   {\ensuremath{\Bz \to \piz\piz}\xspace}
\def\Btopizpiz   {\ensuremath{\Bz \to \piz\piz}\xspace}

\def\Btag {\ensuremath{B_{\rm tag}}}
\def\hpm    {\ensuremath{h^{\pm}}\xspace}
\def\acp {\ensuremath{{\cal A}}\xspace}

\def\alphaeff {\ensuremath{\alpha_{\rm eff}}\xspace}
\def\de {\ensuremath{\Delta E}\xspace}

\def\fish    {\ensuremath{\cal F}\xspace}

\def\da {\ensuremath{\Delta \alpha}\xspace}

\def\Cpizpiz   {\ensuremath{{\cal C}_{\pi^{0}\pi^{0}}}\xspace}
\def\Apipiz   {\ensuremath{{\cal A}_{\pipm\pi^{0}}}\xspace}
\def\Akpiz   {\ensuremath{{\cal A}_{\Kpm\pi^{0}}}\xspace}
\def\Ahpiz   {\ensuremath{{\cal A}_{h^{\pm}\pi^{0}}}\xspace}

\def\Azz       {\mbox{$A_{00}$}\xspace}
\def\Abzz       {\mbox{$\overline{A}_{00}$}\xspace}

\def\Bzz {1.47}
\def\Bzzstat {0.25}
\def\Bzzsyst {0.12}

\def\Czz {-0.49}
\def\Czzstat {0.35}
\def\Czzsyst {0.05}

\def\Bppz {5.02}
\def\Bppzstat {0.46}
\def\Bppzsyst {0.29}

\def\Appz {0.03}
\def\Appzstat {0.08}
\def\Appzsyst {0.01}

\def\Bkpz {13.6}
\def\Bkpzstat {0.6}
\def\Bkpzsyst {0.7}

\def\Akpz {0.030}
\def\Akpzstat {0.039}
\def\Akpzsyst {0.010}

\newcommand{\BABARPubYear}    {07}
\newcommand{\BABARPubNumber}  {039}

\newcommand{\SLACPubNumber} {12682}

\def\figurebox#1#2#3{%
    \def\arg{#3}%
    \ifx\arg\empty
    {\hfill\vbox{\hsize#2\hrule\hbox to #2{\vrule\hfill\vbox to #1{\hsize#2\vfill}\vrule}\hrule}\hfill}%
    \else
    {\hfill\epsfbox{#3}\hfill}%
    \fi}

\long\def\inst#1{\par\nobreak\kern 4pt\nobreak
    {\it #1}\par\vskip 10pt plus 3pt minus 3pt}

\begin{document}

\preprint{\babar-PUB-\BABARPubYear/\BABARPubNumber} 
\preprint{SLAC-PUB-\SLACPubNumber} 
                 
 \begin{flushleft}
 \babar-PUB-\BABARPubYear/\BABARPubNumber\\
 SLAC-PUB-\SLACPubNumber\\
 \end{flushleft}

\title{
{\large \bf
Study of  \Bztopizpiz, \Btopipiz, and \Btokpiz Decays, and 
Isospin Analysis of \Btopipi Decays }
}

%
\author{B.~Aubert}
\author{M.~Bona}
\author{D.~Boutigny}
\author{Y.~Karyotakis}
\author{J.~P.~Lees}
\author{V.~Poireau}
\author{X.~Prudent}
\author{V.~Tisserand}
\author{A.~Zghiche}
\affiliation{Laboratoire de Physique des Particules, IN2P3/CNRS et Universit\'e de Savoie, F-74941 Annecy-Le-Vieux, France }
\author{J.~Garra~Tico}
\author{E.~Grauges}
\affiliation{Universitat de Barcelona, Facultat de Fisica, Departament ECM, E-08028 Barcelona, Spain }
% ...
\collaboration{The \babar\ Collaboration}
\noaffiliation

\date{\today}

\begin{abstract}
We present updated measurements of the branching fractions and \CP
asymmetries for 
\Bztopizpiz, \Btopipiz, and \Btokpiz.
Based on a sample of $383\times 10^{6}$ 
\FourS\to\BB\ decays collected by the \babar\ detector at the \pep2
asymmetric-energy $B$ factory at SLAC, we measure $\BR(\Bztopizpiz)
=(\Bzz \pm \Bzzstat \pm \Bzzsyst) \times 10^{-6}$, ${\BR}(\Btopipiz)
= (\Bppz \pm \Bppzstat \pm \Bppzsyst) \times 10^{-6}$, and
${\BR}(\Btokpiz)   = (\Bkpz \pm \Bkpzstat \pm \Bkpzsyst) \times
10^{-6}$.  We also measure the \CP asymmetries ${\cal C}_{\piz\piz}=\Czz \pm
\Czzstat \pm \Czzsyst $, $ \acp_{\pipm\piz}   = \Appz\pm \Appzstat \pm
\Appzsyst $, and  $\acp_{\Kpm\piz}   = \Akpz \pm \Akpzstat \pm \Akpzsyst$.  
Finally, we present bounds on the CKM
angle $\alpha$ using isospin relations. 
\end{abstract}


\maketitle

In the Standard Model (SM) of particle physics, the charged-current
couplings of the quark 
sector are described by the Cabibbo-Kobayashi-Maskawa (CKM) matrix elements 
$V_{\rm qq^{\prime}}$\cite{ckmref}.  The consistency of multiple
measurements of the sides and angles 
of the CKM Unitarity Triangle provides a stringent test of the SM, and
also provides
constraints on non-SM physics.  The CKM angle $\alpha \equiv$ arg
$[-(V_{\rm td}V_{\rm tb}^{*})/(V_{\rm ud}V^{*}_{\rm ub})]$ can be measured from the 
interference between $b\to u$ quark decays with and without
$\Bz\leftrightarrow\Bzb$ mixing.  In the limit of one (tree)
amplitude, $\sin 2\alpha$ can 
be extracted from the \CP asymmetries in $\Bz\to\pip\pim$ decays
\cite{cpimplied}. However,
the size of the branching fraction of $\Btopizpiz$, relative to
$\Btopipiz$ and $\Bz\to\pip\pim$, indicates that there is another
significant (penguin) 
amplitude, with a different \CP-violating (weak) phase, contributing
to the decay. 
The deviation of the asymmetry obtained from $\Btopipi$ decays,
$\sin 2 \alphaeff$, from $\sin 2\alpha$ can be measured using the
isospin-related decays $\Btopipiz$ and $\Btopizpiz$
~\cite{Isospin,previous,previousBelle}.  
In the SM, the charge  
asymmetry is expected to be very small in the decay \Btopipiz
since penguin diagrams 
cannot contribute to the $I=2$ final state.  However, a non-zero
time-integrated \CP asymmetry in the decay \Btopizpiz is expected if
penguin and tree amplitudes have different weak and \CP-conserving
(strong) phases.

The $\B\to K\pi$ system also exhibits interesting \CP-violating
features, including direct 
\CP violation in $\Bz\to\Kp\pim$ decays \cite{pipiResults,pipiResultsII}.
 Sum rules derived from U-spin symmetry and parameters from the $B\to\pi\pi$
system relate the branching fraction and charge asymmetry of
$\Btokpiz$ decays to other decays in the $K\pi$ system~\cite{SumRule,SU3}.
 The \CP asymmetry in \btokpiz is expected to have the same sign and
roughly the same magnitude as the \CP asymmetry in $\Bz\to\Kp\pim$ in
the absence of color-suppressed tree and electroweak-penguin amplitudes.

Based on a sample of $383\times 10^{6}$ \FourS\to\BB decays, we report
updated measurements of the branching fraction for $\Btopizpiz$
and the time-integrated \CP asymmetry, 
${\cal C}_{\piz\piz}$, defined as
\begin{equation}
{\cal C}_{\piz\piz} \equiv \cfrac{|\Azz|^2 - |\Abzz|^2}{|\Azz|^2 + |\Abzz|^2}\,  
\end{equation}
where $\Azz(\Abzz)$ is the $\Bz(\Bzb)\to\piz\piz$ decay amplitude.
We also measure the branching fractions for $\Btohpiz$ 
($h^{\pm} = \pipm,\Kpm$) and the corresponding 
charge asymmetries
\begin{equation}
{\cal A}_{h^{\pm}\piz} \equiv \cfrac{
N_{\pim\piz} - N_{\pip\piz}}
{N_{\pim\piz} + N_{\pip\piz}}\, 
\end{equation}
where $A_{\pm 0}(\overline{A}_{\pm 0})$ is the \Bp(\Bm) decay amplitude.

The \babar\ detector is described in Ref.~\cite{babarnim}.
Charged particle momenta are measured with a
tracking system consisting of a five-layer silicon
vertex tracker (SVT) and a 40-layer drift chamber (DCH) surrounded
by a $1.5$-T solenoidal magnet.  An electromagnetic calorimeter
(EMC) comprising $6580$ CsI(Tl) crystals is used to measure 
photon energies and positions. The photon energy resolution
in the EMC is  
$\sigma_{E}/E = \left\{2.3 (\gev)^{1/4} / E^{1/4} \oplus 1.9 \right\} \%$,
and the angular resolution from the interaction point is  
$\sigma_{\theta} = 3.9^{\rm  o}/\sqrt{E/\gev}$. Charged hadrons are
identified with a detector 
of internally reflected Cherenkov light (DIRC) and ionization
measurements in the tracking detectors.  The average $K$-$\pi$
separation in the DIRC varies from $12\sigma$ at a laboratory momentum
of $1.5~\gevc$ to $2\sigma$ at $4.5~\gevc$.

For the reconstruction of \Btohpiz events, we require the track from
the $B$ candidate to
have at least 12 hits in the DCH and be
associated with at least 5 photons in the DIRC.  The measured
Cherenkov opening angle $\theta_C$ must be within $4\sigma$ of the
expectation for the pion or kaon
hypothesis and  $\theta_C$ must be greater than  10\mrad from the
proton hypothesis.  Electrons are removed from the sample by vetoing
candidates based on their energy loss in the SVT and DCH and a comparison of the
track momentum and deposited energy in the EMC.

While \piz meson candidates are mostly formed from two EMC
clusters, we 
increase our \piz efficiency compared to Ref. \cite{previous} by
$\sim10\%$ by including \piz  
candidates consisting of two overlapping photon clusters (``merged'' \piz)
and candidates with one photon cluster and two tracks
consistent with being a photon conversion inside the detector.
Photon conversions are selected from pairs of oppositely 
charged tracks with an invariant mass less than 30 \mevcc, a vertex
that lies within the detector, and a total momentum
vector that points back to the beamspot.
EMC clusters are required to have energies greater than 0.03 \gev and a
transverse shower shape consistent with a photon.  
To reduce the background from random photon combinations, the cosine
of the angle between the direction of the decay photons 
in the center-of-mass system of the parent \piz and the \piz  flight
direction in the lab frame must be less than 0.95. For candidates
consisting of two EMC clusters or one cluster and a converted photon,
the reconstructed 
\piz mass is required to be between 110 and 160 \mevcc, and the
candidates are then kinematically fit with their mass constrained
to the \piz mass. We
distinguish merged \piz candidates from single photons and other
neutral hadrons using the second transverse moment, 
$S = \sum_{i} E_{i} \times (\Delta\alpha_i)^{2}/ E$, where $E_{i}$ is
the energy deposited in each CsI(Tl) crystal, and 
$\Delta \alpha_{i}$ is the angle between the cluster centroid and
the crystal.  Because merged $\piz$s are caused by two
overlapping photon clusters, they have a larger $S$ than
solitary photons.
  We use a large sample of $\piz$s from
$\tau^{\pm}\to\rho^{\pm}\nu$ decays to validate that our Monte Carlo
simulation (MC) accurately
simulates merged $\piz$s and photon conversions, as well as our
overall \piz efficiency.  

We use two kinematic variables to isolate \Btopizpiz and \Btohpiz
candidates from the large background of \epem\to\qqbar $(q=u,d,s,c)$
continuum events.  The first is the beam-energy-substituted mass $\mes = \sqrt{
  (s/2 + {\bf p}_{i}\cdot{\bf p}_{B})^{2}/E_{i}^{2}- {\bf
    p}^{2}_{B}}$, where $\sqrt{s}$ is the total \epem center-of-mass (CM)
energy,  $(E_{i},{\bf p}_{i})$ is the four-momentum of the initial
\epem system, and ${\bf p}_{B}$ is the \B-candidate momentum, both measured in
the laboratory frame.  The second variable is \de
$ = E_{B} - \sqrt{s}/2$, where $E_{B}$ is the \B candidate energy in
the CM frame.  For \Btohpiz, we require $\mes>5.22\gevcc$ and 
$-0.11\gev < \de < 0.15\gev$. We define the main signal region in the
\Btopizpiz analysis as $\mes>5.20\gevcc$ and 
$|\de| < 0.20\gev$.  

To further discriminate the signal from \qqbar
backgrounds, we exploit the event topology variable $\theta_S$: the
angle in the CM frame between the sphericity axis of the $B$
candidate's decay products 
and that of the remaining neutral clusters and charged tracks in the
rest of the event.  Since the distribution of
$|\cos\theta_S|$ peaks at 
$1$ for \qqbar events, we require $|\cos\theta_S|<0.8$ $(0.7)$
for events with a \Btohpiz (\Btopizpiz) candidate.  
To further improve background separation,
we construct a Fisher discriminant ${\cal F}$ from the sums $\sum_i p_i$ and
$\sum_i p_i \cos^2{\theta_i}$, where $p_{i}$ is 
the CM momentum and $\theta_{i}$ is the angle with respect to the thrust
axis of the \B candidate's daughters, in the CM frame, of all tracks and
clusters not used to reconstruct the \B meson.

We use an extended, unbinned maximum likelihood (ML) fit to determine the
number of signal events and the associated asymmetries. The
probability density function (PDF) 
${\cal P}_i\left(\vec{x}_j;  \vec{\alpha}_i\right)$ for event $j$ and signal or
background hypothesis $i$ is the product of PDFs for the variables
$\vec{x}_j$, given the set of parameters  $\vec{\alpha}_i$.  The
likelihood function ${\cal L}$ is
\begin{equation}
{\cal L}= \exp\left(-\sum_{i=1}^M n_i\right)\,
\prod_{j=1}^N \left[\sum_{i=1}^M n_i {\cal P}_i\left(\vec{x}_j;
\vec{\alpha}_i\right)
\right]\, ,
\end{equation}
where $N$ is the number of events, $n_i$ is the PDF coefficient for
hypothesis $i$, and $M$ is
the total number of signal and background hypotheses.  
 
In the \Btopizpiz fit, the variables $\vec{x}_j$ are \mes, \de, and \fish.  
In addition to the signal and \qqbar background, we expect background events
from the charmless decays $\Bpm\to\rho^{\pm}\piz$ 
and \Bz\to\KS\piz (\KS\to\piz\piz) to contribute $61\pm 7$ events in
the signal region, as
determined from MC, so we include an additional component
in the fit to account for this \BB background.   For
the \Btopizpiz signal and the \BB background, we
observe a correlation coefficient between \mes and \de of $\sim0.2$, so a
two-dimensional PDF, derived from MC simulation, is used to
parameterize these distributions. The \qqbar background PDF is described
by an ARGUS threshold function ~\cite{Argus} in \mes and a polynomial
in \de. We divide the \fish distribution from signal MC into ten
equally-populated bins, and use a parametric step function to
describe the distribution for all of the signal and background hypotheses.
We fix the relative size of the \fish bins for the signal and \BB
background to values taken from MC. These values are verified with a sample
of fully reconstructed $B$ meson decays.  
Continuum \fish parameters are free in the fit.

In order to measure the time-integrated \CP asymmetry 
${\cal C}_{\piz\piz}$, we use the remaining 
tracks and clusters in a multivariate technique \cite{sin2betaPRD02}
to determine the 
flavor (\Bz or \Bzb) of the other $B$ meson in 
the event ($\Btag$).
Events are assigned to
one of seven mutually exclusive categories $k$ (including untagged
events with no flavor information) based on the estimated mistag probability $w_{k}$ and on
the source of 
the tagging information.   The PDF
coefficient for \Btopizpiz is given by
\begin{equation}
n_{\piz\piz, k} = \frac{1}{2} f_{k} N_{\piz\piz} \Bigl[ 1 - s_j
  (1-2\chi_{\rm d})(1-2w_k) {\cal C}_{\piz\piz} \Bigr],
\end{equation}
where $N_{\piz\piz}$ is the total number of \Btopizpiz decays,  
$\chi_{\rm d} = 0.188 \pm 0.004$~\cite{pdg} is the time-integrated mixing
probability, and $s_j=+1(-1)$ 
when the \Btag\ is a \Bz (\Bzb). The fraction of events in each
category, $f_{k}$, and the mistag rate  are determined from a
large sample of $\Bz\to D^{(*)} (n \pi) \pi$ decays.  

For the \Btohpiz fit, along with \mes, \de, and \fish, we include the
Cherenkov angle $\theta_C$ to measure 
the \Btopipiz and \Btokpiz yields and asymmetries simultaneously.  The
difference between the expected and measured Cherenkov angle, divided
by the uncertainty, is described by two Gaussian distributions.  The
values for \mes and \de are calculated assuming the track is a
pion, so a \Btokpiz event will have \de shifted
by a value dependent on the track momentum, typically $-45\mev$.  For the
signal, the \mes and \de distributions are modeled as  Gaussian
functions with  low-side power-law tails. The means of these
distributions and the \mes width are determined in the fit, while the
\de width
is determined by MC simulation.  We expect $69\pm3$ background events in the
\Btopipiz signal region from other $B$ meson decays, mainly from the
same $B$ decays as in the \Btopizpiz case.  For the \Btokpiz signal
region we expect $9\pm2$ events from $B\to X_{s}\gamma$ and
$\Bz\to\rho^{+}\Km$. The PDFs for the \BB backgrounds, the
\qqbar background, and the signal \fish are all treated the same as in
the \Btopizpiz case.   The PDF
coefficient for \btohpiz is given by 
$n_i = \frac{1}{2}N_i \left(1-q_j{\cal A}_i\right)$,
where ${\cal A}_i$ is the charge asymmetry, and $q_j = \pm 1$ is the charge of the
$B$ candidate. 

\begin{figure}[!tbph]
\begin{center}
  \includegraphics[width=0.49\linewidth]{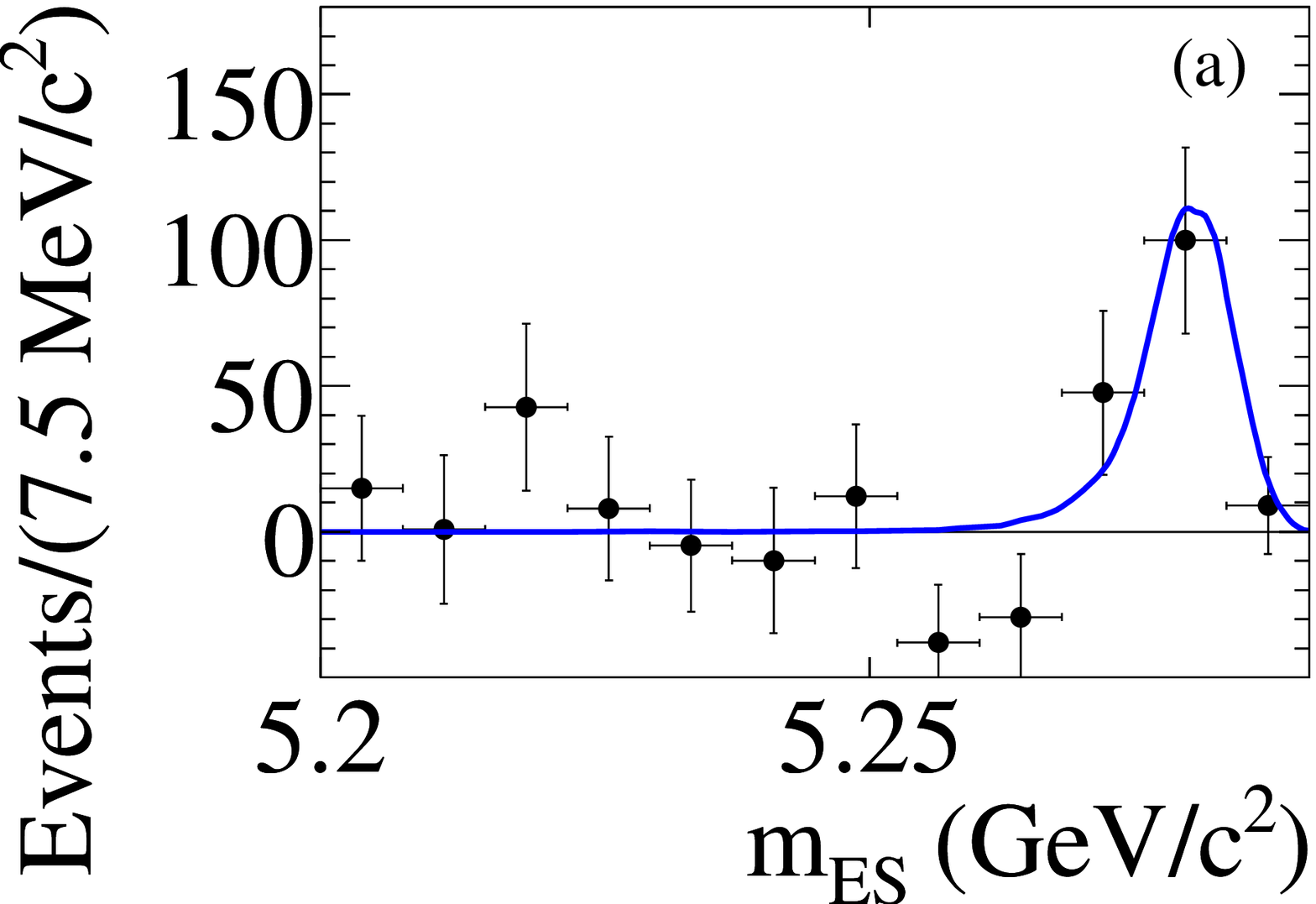}
  \includegraphics[width=0.49\linewidth]{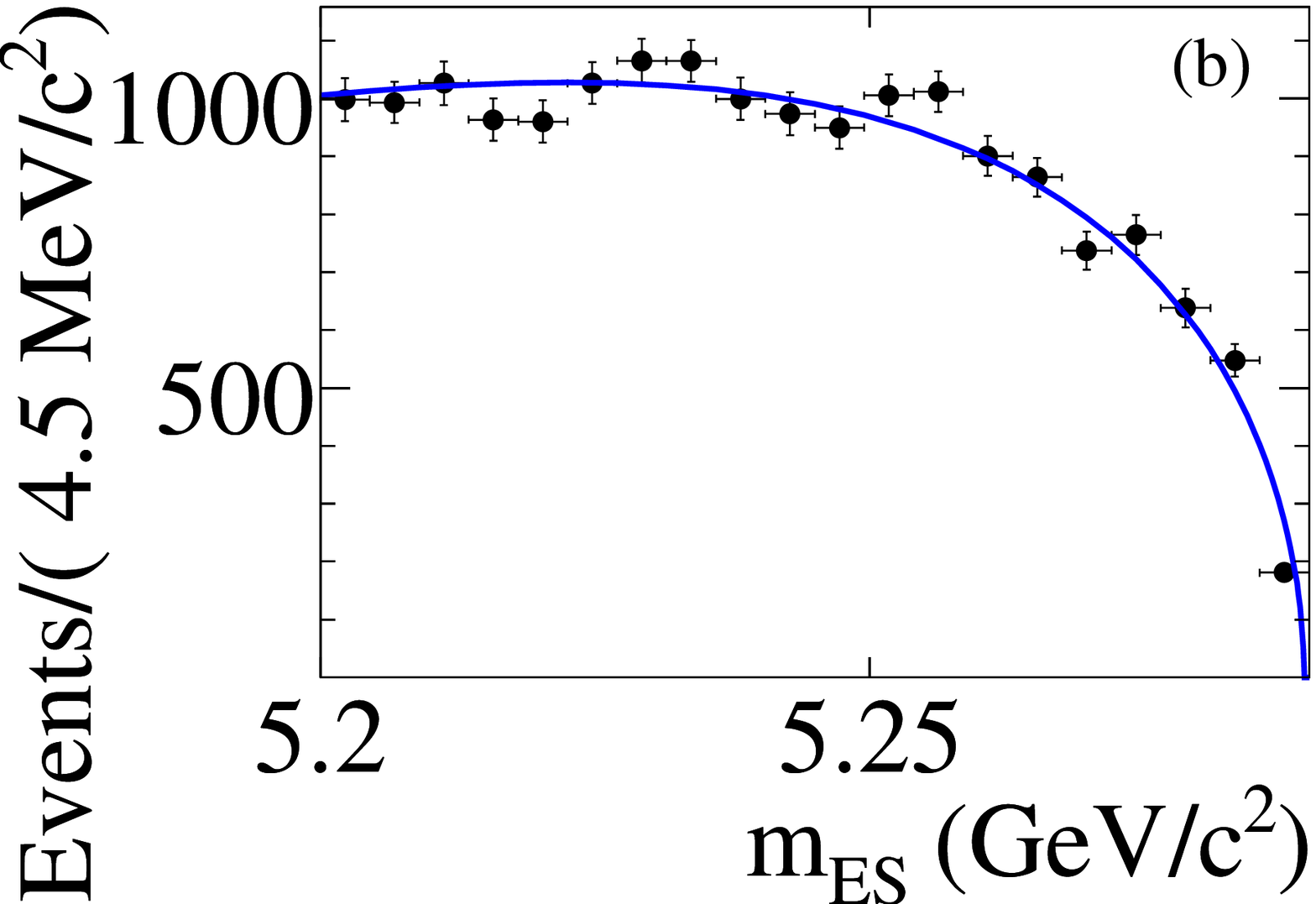}
  \includegraphics[width=0.49\linewidth]{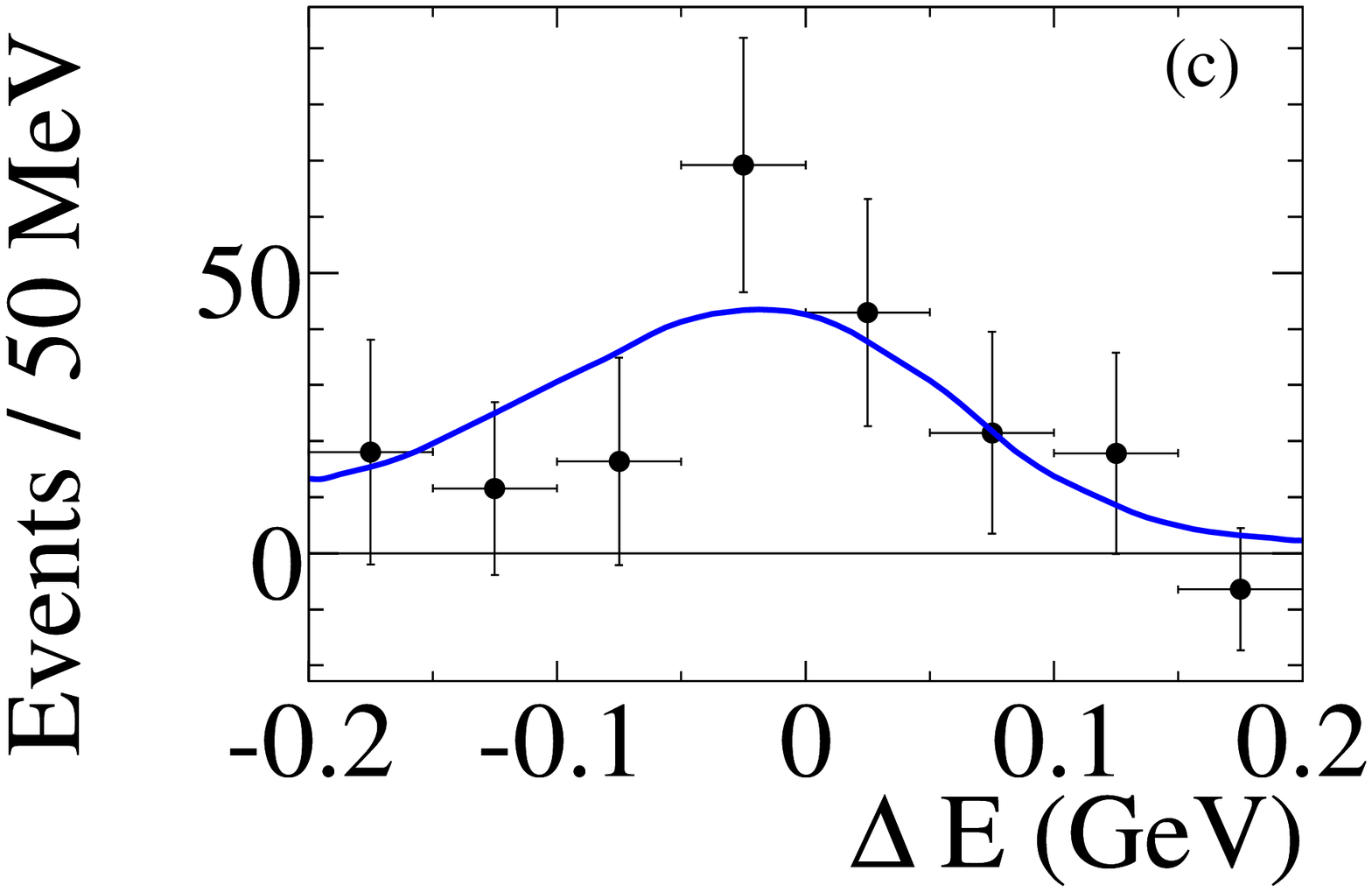}
  \includegraphics[width=0.49\linewidth]{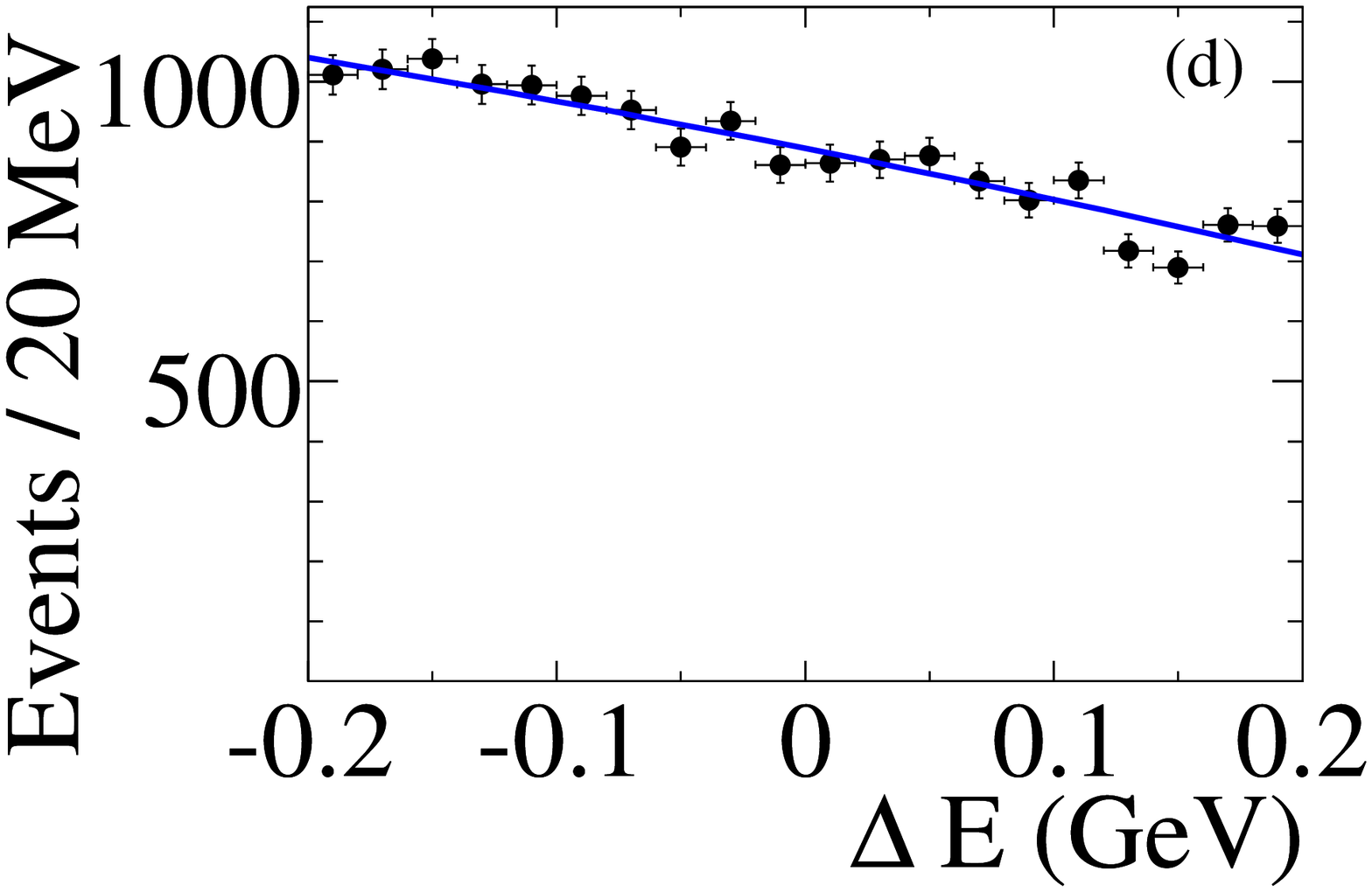}
  \includegraphics[width=0.49\linewidth]{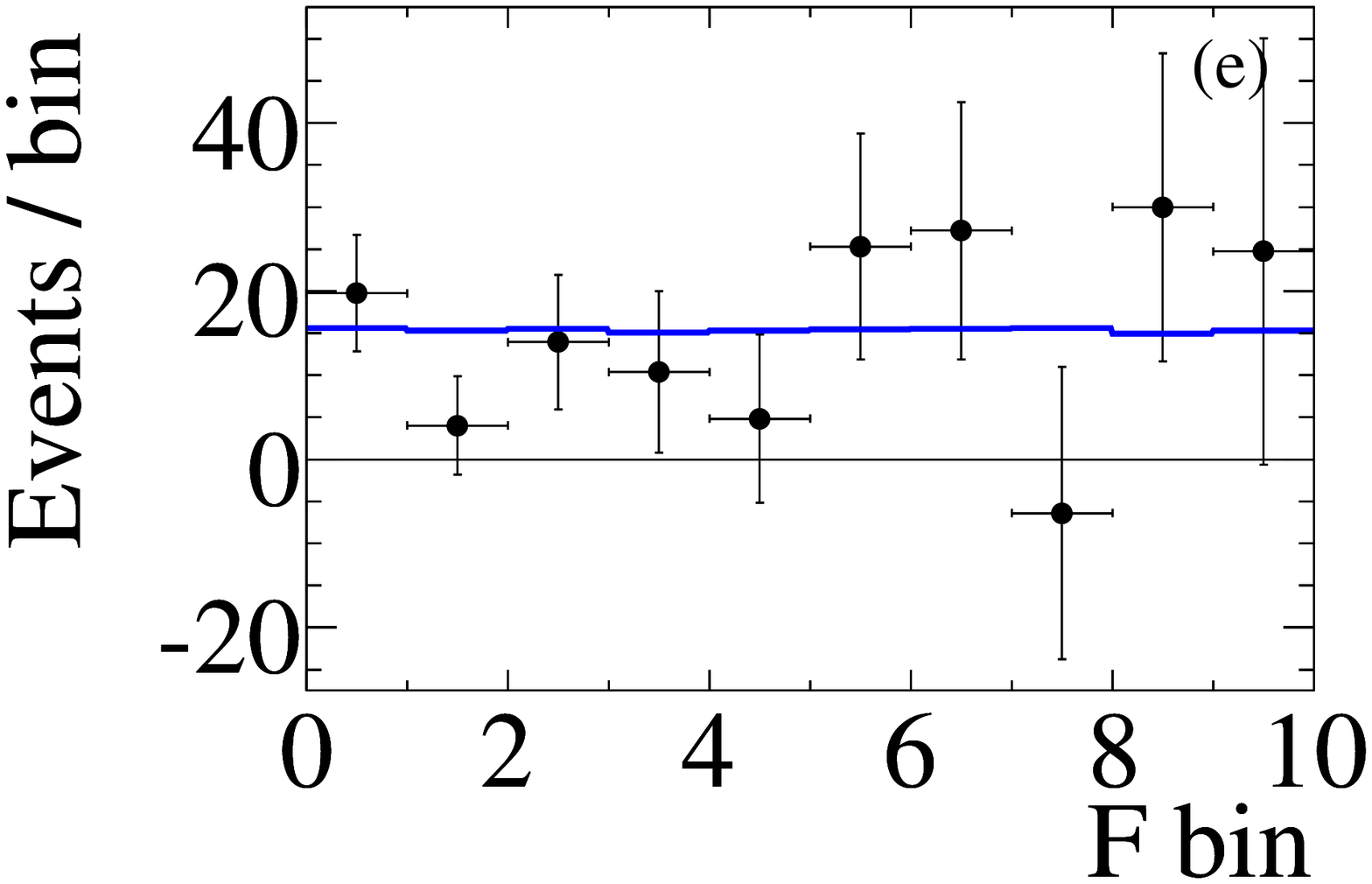}
  \includegraphics[width=0.49\linewidth]{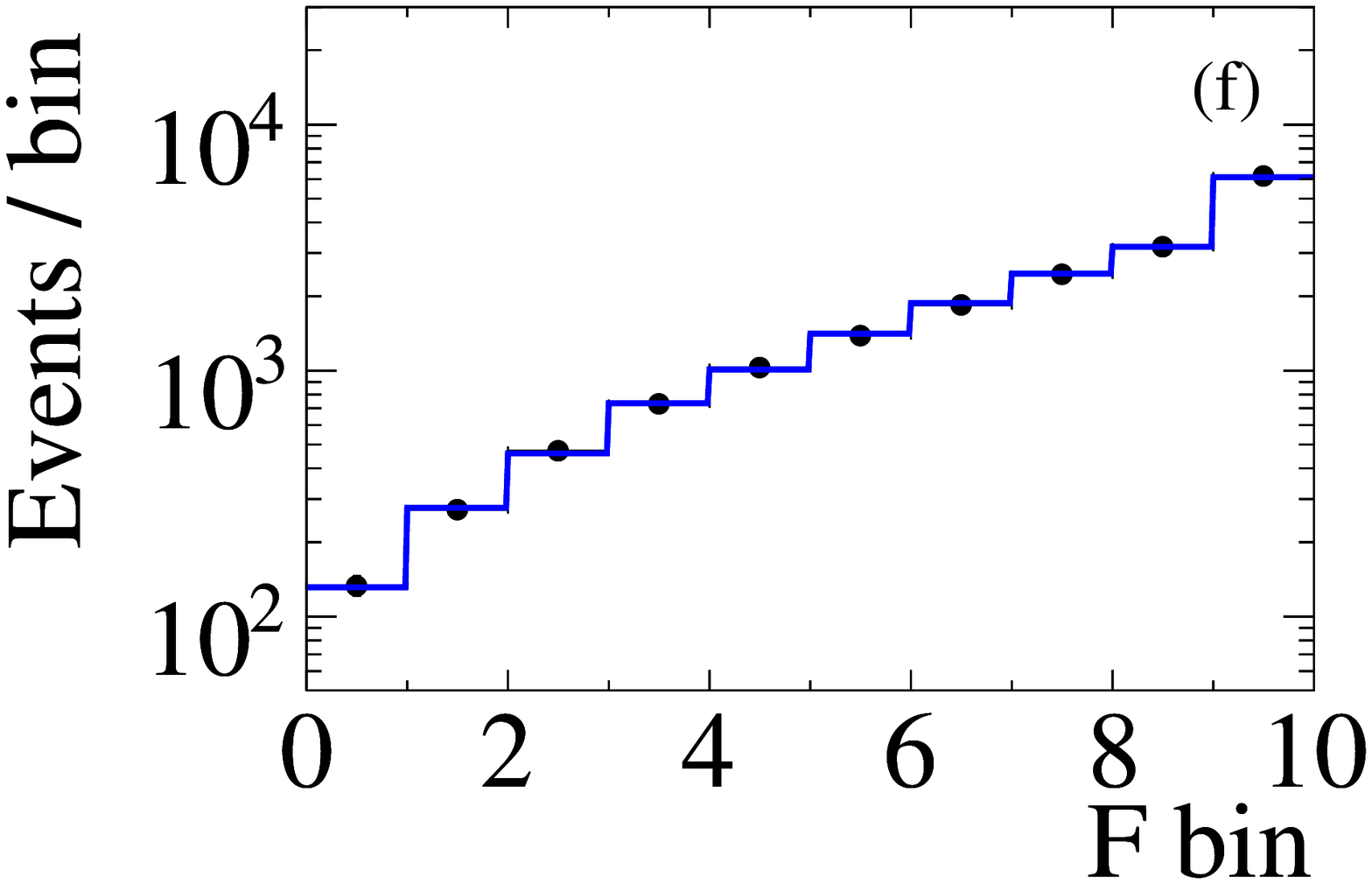}
  \caption{Distributions made with the event weighting
    and background 
    subtraction method described in Ref.~\cite{sPlots} and PDF projections for
    the likelihood fit variables in the \Btopizpiz fit.
    Shown are \mes (a,b), \de (c,d) and \fish (e,f) for signal
    (a,c,e) and continuum background (b,d,f). }
  \label{fig:pizpizplots}
\end{center}
\end{figure}

\begin{figure}[!tbph]
\begin{center}
  \includegraphics[width=0.49\linewidth]{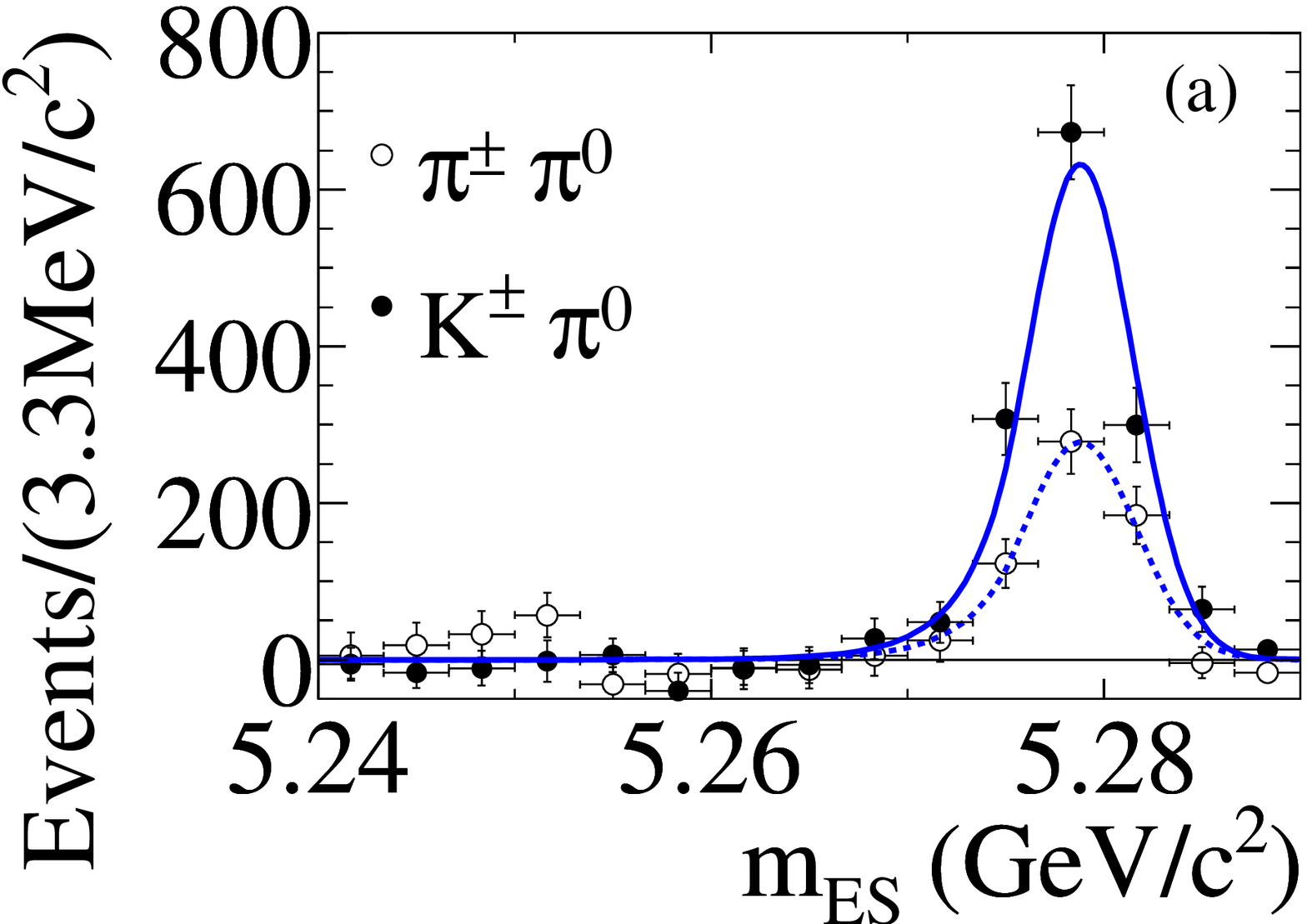}
  \includegraphics[width=0.49\linewidth]{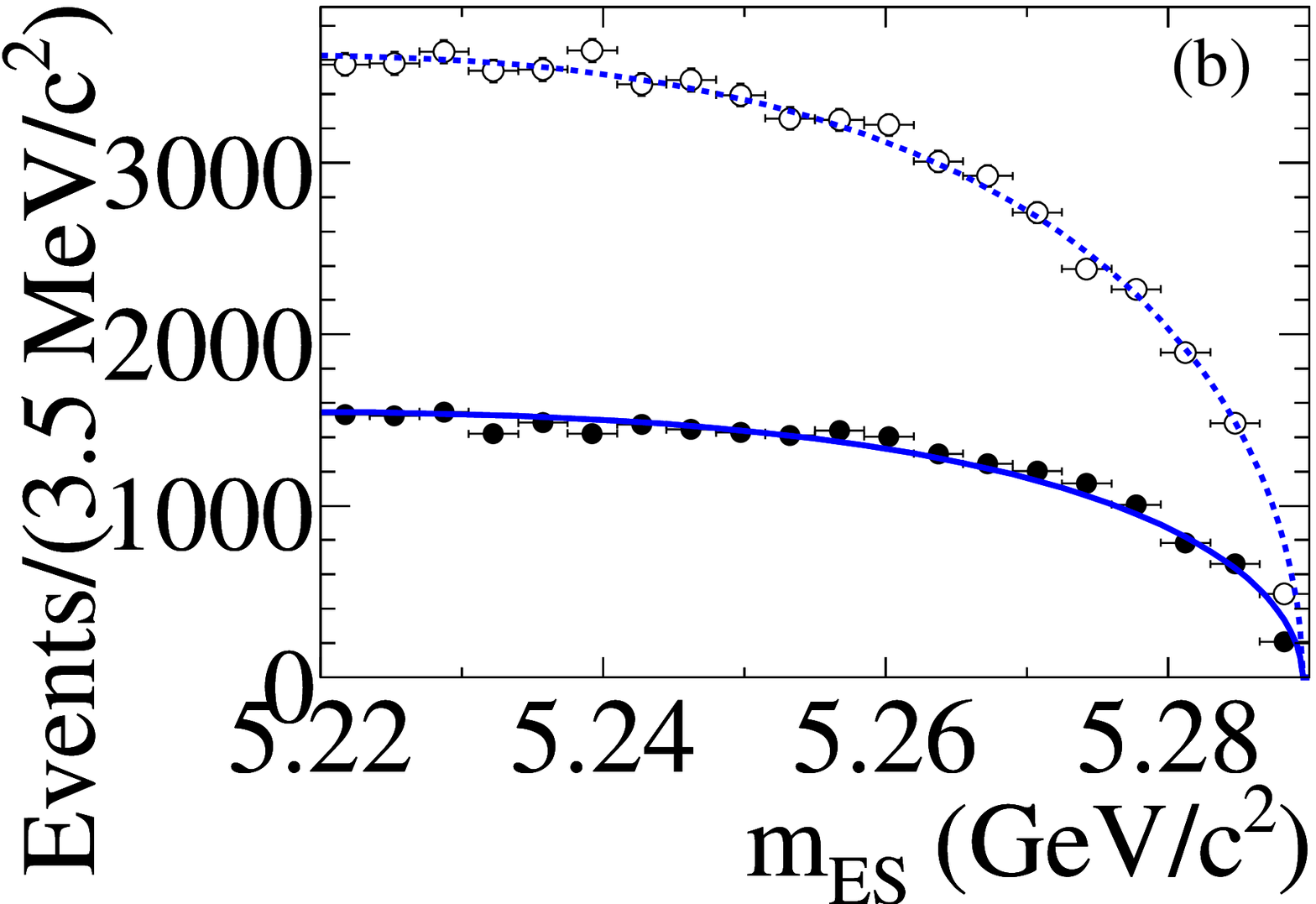}
  \includegraphics[width=0.49\linewidth]{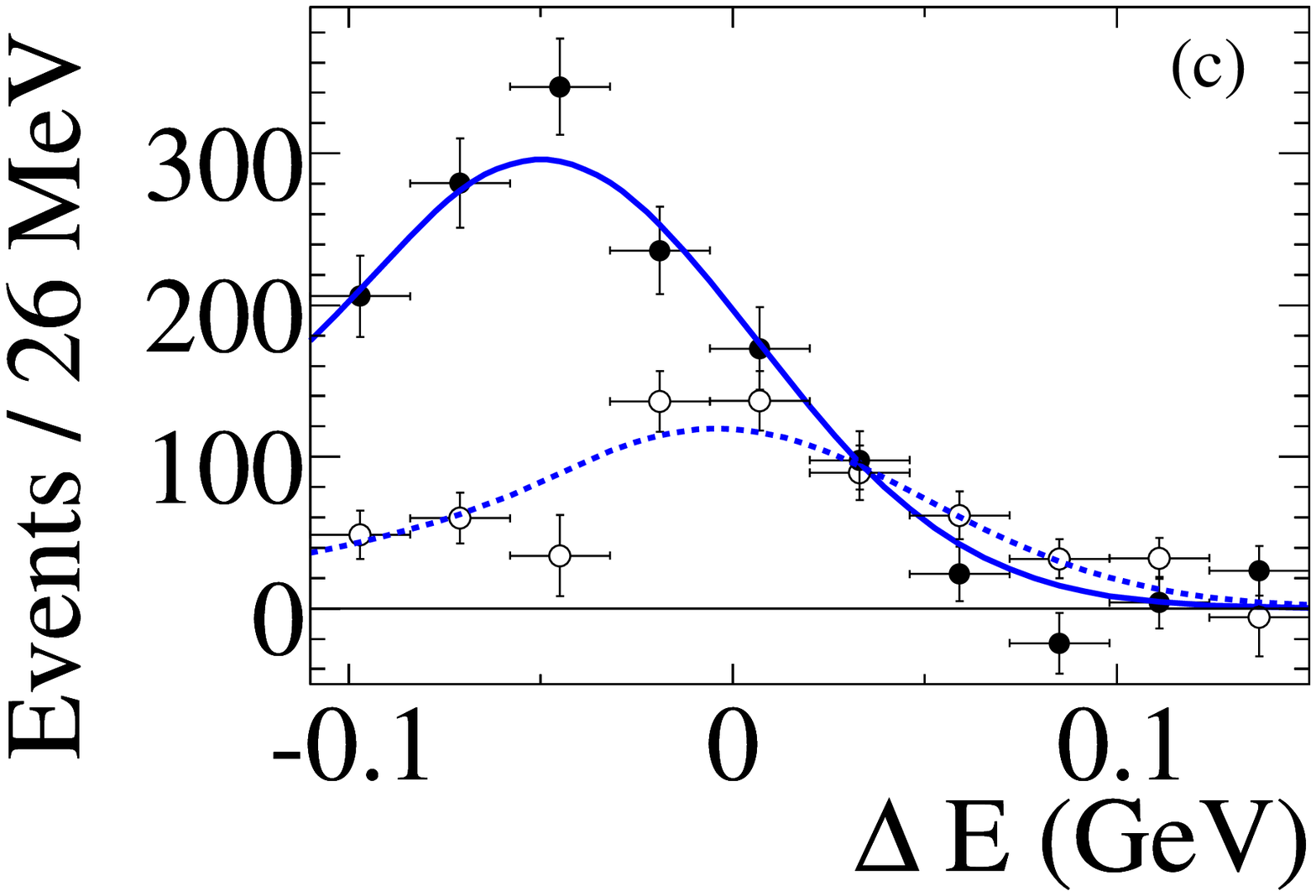}
  \includegraphics[width=0.49\linewidth]{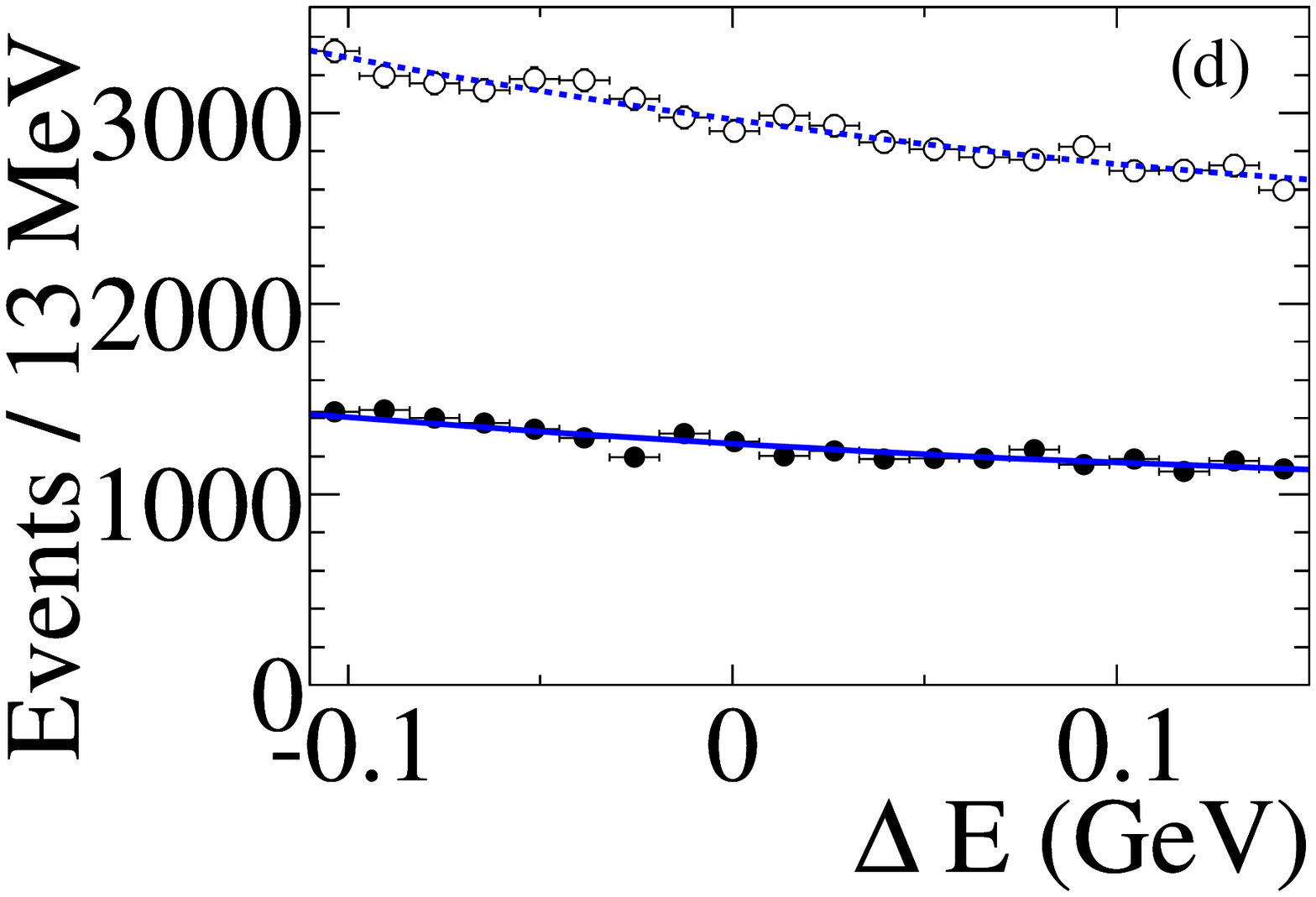}
  \includegraphics[width=0.49\linewidth]{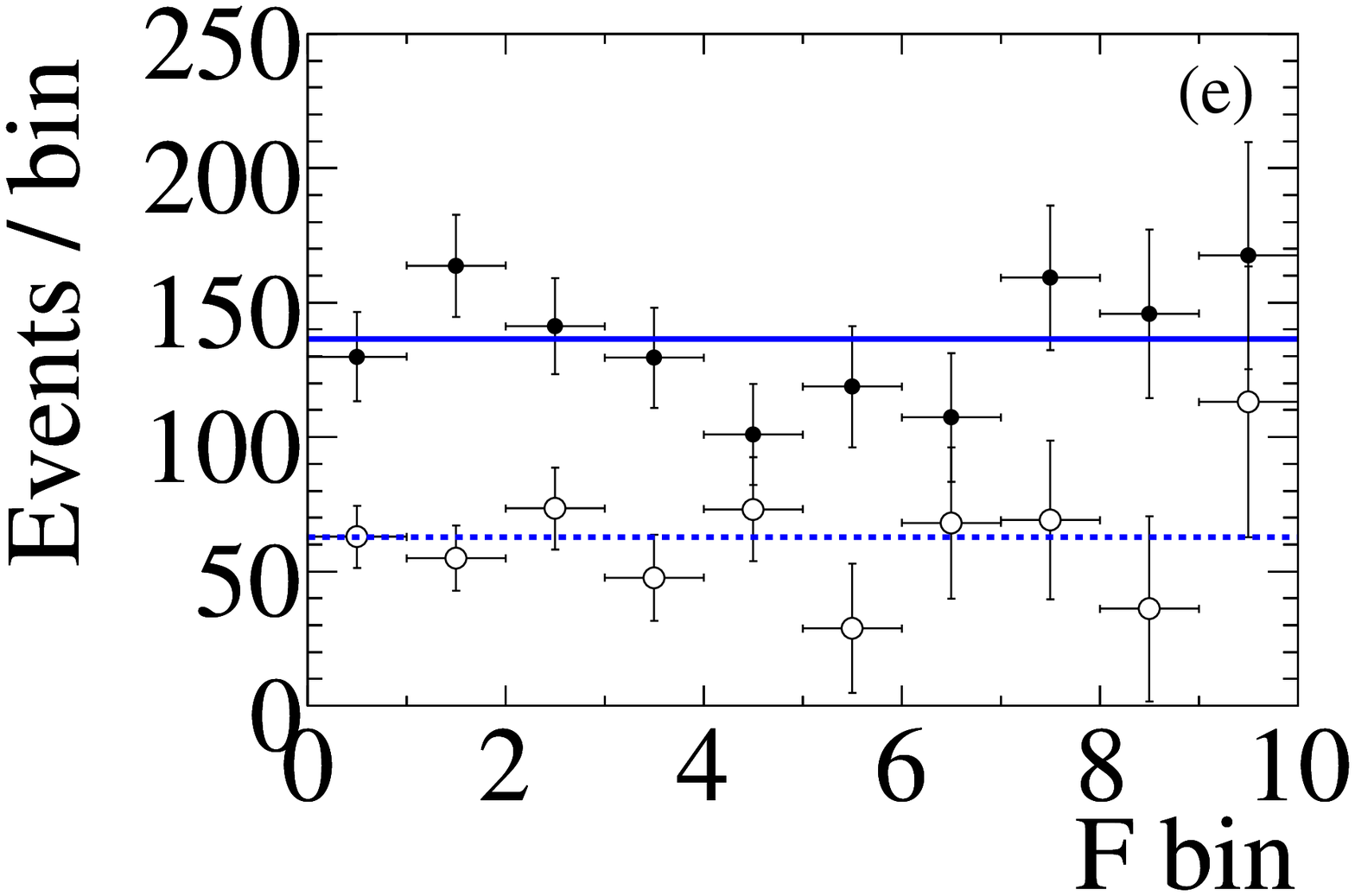}
  \includegraphics[width=0.49\linewidth]{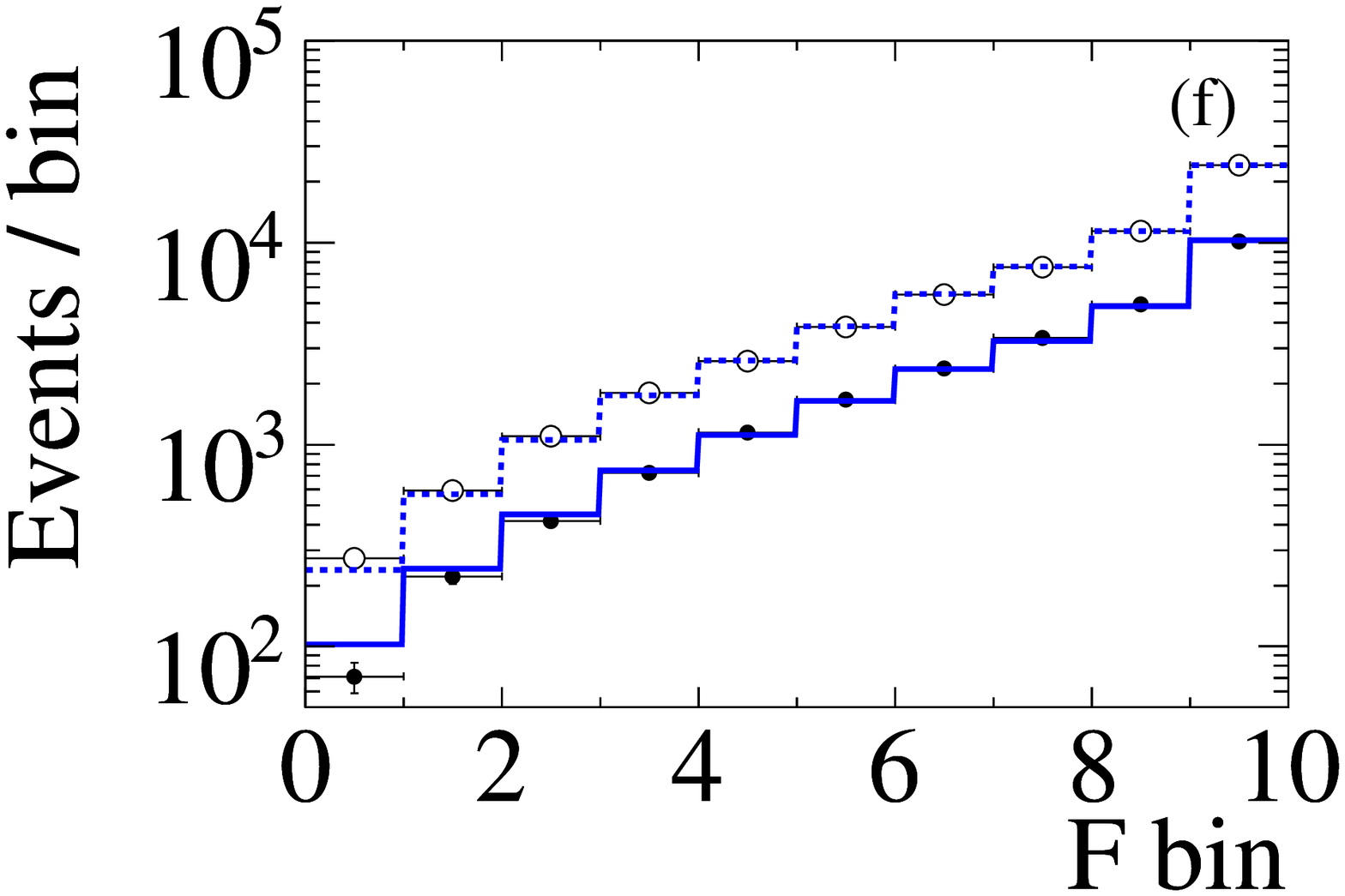}
  \caption{Distributions made with the event weighting and background
    subtraction method described in Ref.~\cite{sPlots} and PDF projections of
    the likelihood fit variables from the \btohpiz fit.
    Shown are \mes (a,b), \de (c,d) and \fish (e,f)
    distributions for signal 
    (a,c,e) and continuum background (b,d,f). 
     PDF projections for the \btokpiz signal and background are shown
     as solid  lines, while the PDF projections for the
    \btopipiz signal and background are shown as dashed lines.
    \btopipiz signal and background is
    shown as open circles while the \btokpiz is shown
    as solid circles.}
  \label{fig:hpi0Plots}
\end{center}
\end{figure}

The results from the \Btopizpiz and \Btohpiz  ML fits are summarized in
Table \ref{tab:results}. In a total of 17,881 events we find
$154\pm27$ \Btopizpiz  decays and an 
asymmetry ${\cal C}_{\piz\piz} = \Czz \pm \Czzstat$.
For the \btohpiz fit, we find $627\pm58$ \btopipiz and
$1364\pm57$ \btokpiz events in a total of 85,895 events.  
All of the correlations among the signal variables are less than
$5\%$. 
In Fig.~\ref{fig:pizpizplots} we use the event weighting  and
background subtraction method described 
in Ref.~\cite{sPlots} to show signal and background distributions for
\Btopizpiz events. Signal and background distributions for \btohpiz
events are shown in Fig.~\ref{fig:hpi0Plots} using the same method.

\begin{table*}[!btph]
  \begin{center}
    \caption{ The results for the \Bztopizpiz and \Btohpiz decays.
      For each mode we show the number of signal events, $N_S$, 
      number of continuum events, $N_{\rm cont}$, number of $B$-background
      events, $N_{\rm Bbkg}$, total detection efficiency $\varepsilon$,
      branching fraction \BR, and asymmetry ${\cal A}_{h^{\pm}\piz}$
      or ${\cal C}_{\piz\piz}$.  Uncertainties are statistical for
      $N_{S}$ and $N_{\rm cont}$, while for the branching fractions and
      asymmetries they are statistical and systematic, respectively.} 
    \label{tab:results}
    \begin{tabular}{l|ccccccccccc}\hline\hline
      Mode        &  $N_{S}$      &&$N_{\rm cont}(10^3)$&& $N_{\rm Bbkg}$ && $\varepsilon$ (\%) && \BR($10^{-6}$)      & & Asymmetry \\\hline
      \Bztopizpiz &  $154\pm 27 $ &&$17.67\pm0.13$  && $61\pm7$ && $27.3$ &&  $\Bzz \pm\Bzzstat\pm\Bzzsyst  $ && $\Czz\pm\Czzstat\pm\Czzsyst$ \\
      \Btopipiz   &  $627\pm 58 $ &&$58.75\pm0.24$  && $69\pm3$ && $32.5$ &&  $\Bppz\pm\Bppzstat\pm\Bppzsyst$ && $\Appz\pm\Appzstat\pm\Appzsyst$\\
      \Btokpiz    &  $1364\pm 57$ &&$25.07\pm0.17$  && $9\pm2$  && $26.6$ &&  $\Bkpz\pm\Bkpzstat\pm\Bkpzsyst$ && $\Akpz\pm\Akpzstat\pm\Akpzsyst$\\\hline\hline
    \end{tabular}
  \end{center}
\end{table*}

In order to account for a small bias in the \btohpiz asymmetries
arising from the difference in the \pip and \pim
reconstruction efficiencies and the \Kp and \Km hadronic interaction
cross-sections in the \babar\ detector, the 
\btopipiz asymmetry is corrected by $+0.005\pm0.004$ and the \btokpiz
asymmetry is corrected by $+0.008\pm0.008$.  We determine the $\pipm\piz$
bias from a study of $\tau^{\pm}\to\rho^{\pm}\nu$ decays and
verify it using the continuum background in data.  For the \btokpiz
charge asymmetry bias, we use the continuum background and combine
the results of the \pipm\piz asymmetry study and the \Kpm\pimp
asymmetry study in Ref. \cite{pipiResults}.   After the bias
correction we find ${\cal A}_{\pipm\piz} = \Appz\pm\Appzstat$ and
${\cal A}_{\Kpm\piz} = \Akpz\pm\Akpzstat$.

We evaluate the systematic errors on the branching fractions and
asymmetries either using data control samples or by varying
fixed parameters and refitting.  The systematic uncertainties on the
branching fraction and asymmetry measurements are summarized in Tables
\ref{tab:BRSystErrs} and \ref{tab:AsymSystErrs}, respectively.
The largest systematic errors for the \Btopizpiz and \Btohpiz branching
fractions are 
from uncertainties in the \piz reconstruction efficiency, signal selection
efficiencies, \fish parameters, and \BB background
yields.  We simulate
radiative effects using the PHOTOS simulation package \cite{photos}
and assign a systematic error equal to 
the difference between PHOTOS and the scalar QED calculation in
Ref.~\cite{ebgi}. For the \Btohpiz analysis, we also include as a
systematic a small ($<2\%$) fit 
bias due to correlation among fit variables.  The largest systematic
uncertainties in the measurement 
of \Cpizpiz are from the uncertainty on the $B$ background \CP
content, tag-side interference,
and the tagging fractions and asymmetry of $\Btag$.   The major 
contributions to the systematic error on \Ahpiz are from the detector
charge asymmetry and the \de and \fish PDF parameterization.

\begin{table}[htbp]
  \begin{center}
    \caption{Systematic errors on the branching fractions for
      \btopipiz, \btokpiz, and \Btopizpiz. }
    \begin{tabular} {l|ccc}\hline\hline
      Source         & $\Delta$$\BR(\pipm\piz)$ &$\Delta$\BR($\Kpm\piz$)&$\Delta$\BR($\piz\piz$) \\\hline
      \piz efficiency      & $3.0\%$ & $3.0\%$ & $6.0\%$ \\
      \mes and \de PDF     & $1.7\%$ & $1.7\%$ & $4.0\%$ \\
      Selection efficiency & $2.8\%$ & $3.0\%$ & $2.7\%$ \\
      \fish PDF            & $2.5\%$ & $0.7\%$ & $1.7\%$ \\
      \BB backgrounds        & $0.2\%$ &$<0.1\%$ & $2.1\%$ \\
      PHOTOS               & $1.9\%$ & $1.1\%$ & ---     \\
      Fit bias             & $1.7\%$ & $1.2\%$ & ---     \\
      Luminosity           & $1.1\%$ & $1.1\%$ & $1.1\%$ \\
      Tracking efficiency  & $0.5\%$ & $0.5\%$ & ---     \\\hline
      Total                & $5.8\%$ & $5.0\%$ & $8.2\%$\\\hline\hline
    \end{tabular}
    \label{tab:BRSystErrs}
  \end{center}
\end{table}

\begin{table}[htbp]
  \begin{center}
    \caption{A summary of the systematic errors on the asymmetries
      \Apipiz, \Akpiz, and \Cpizpiz. All values are expressed in units
      of $10^{-2}$. }
    \begin{tabular} {l|ccc}\hline\hline
      Source         & $\Delta$(\Apipiz) &$\Delta$(\Akpiz)&$\Delta$(\Cpizpiz) \\\hline
      \BB backgrounds        & $0.2 $ &$<0.1 $ & $3.4 $ \\
      Tagging              & ---    & ---    & $2.5 $ \\
      Tag-side interference& ---    & ---    & $1.6 $ \\
      PDF parameters       & $0.8 $ & $0.6 $ & $0.9 $ \\
      Detector asymmetry   & $0.4 $ & $0.8 $ & ---     \\
      Measured $\chi_{\rm d}$ error & ---    & ---    & $0.8 $ \\\hline
      Total                & $0.9 $ & $1.0 $ & $4.7 $\\\hline\hline
    \end{tabular}
    \label{tab:AsymSystErrs}
  \end{center}
\end{table}

\begin{figure}[!tbph]
\begin{center}
  \includegraphics[width=0.8\linewidth]{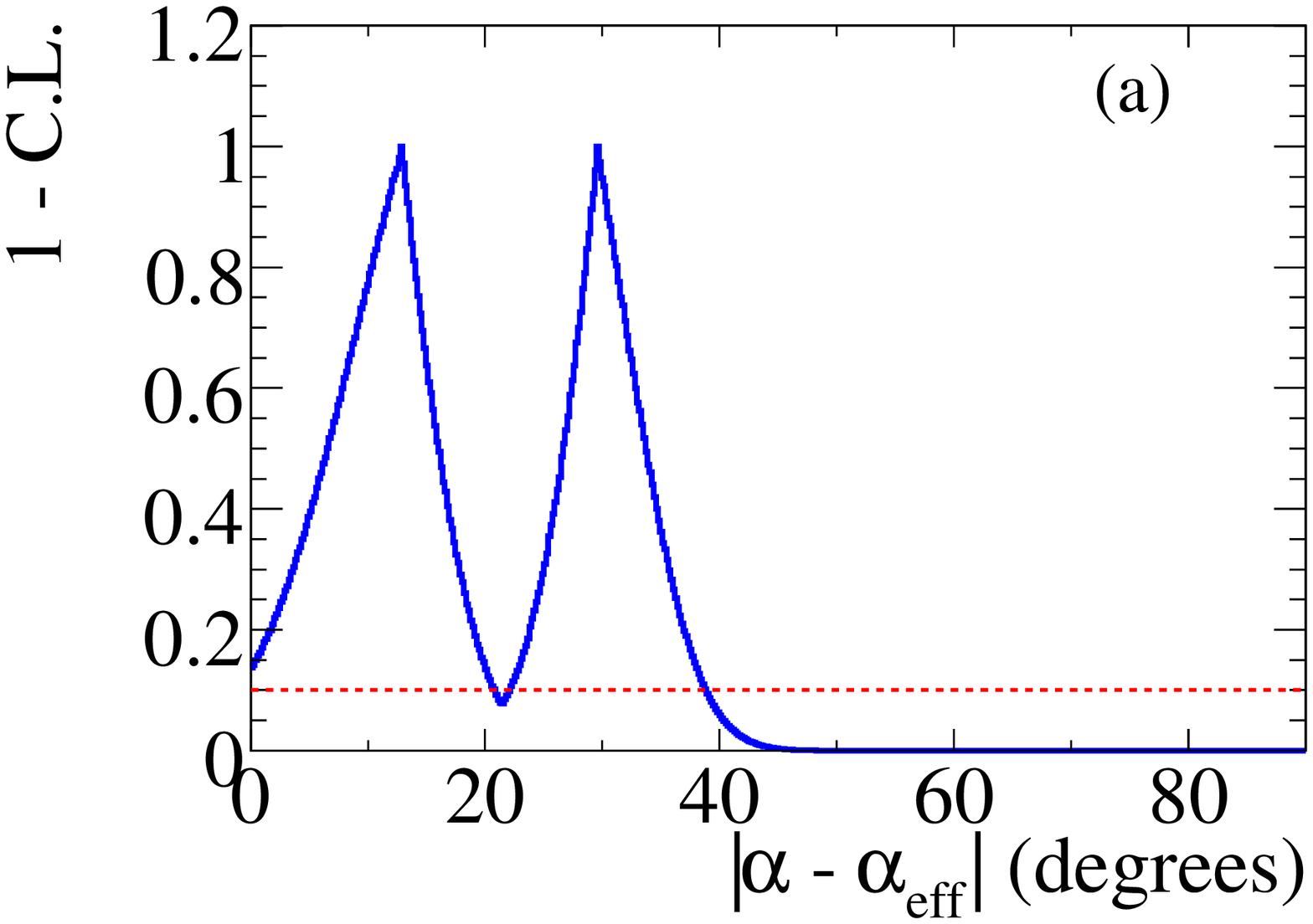}
  \includegraphics[width=0.8\linewidth]{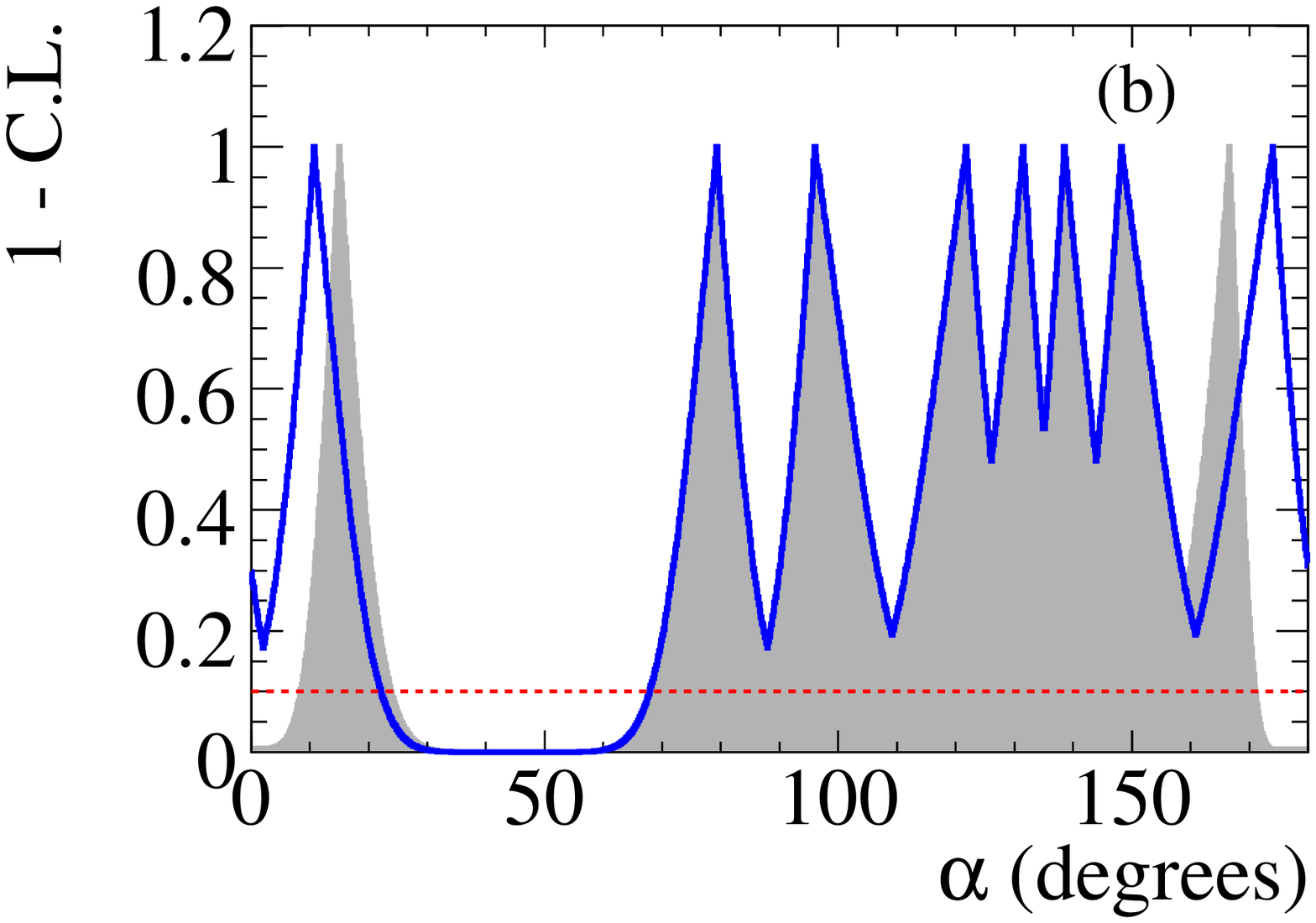}
  \caption{Constraints on (a) the angle $\da = \left| \alpha -
      \alphaeff\right|$ and (b) $\alpha$ expressed as
    one minus the confidence level as a function of angle.  We find an
    upper bound on \da of $39\degrees$ at the $90\%$ confidence
    level. In (b) the curve shows the bounds on $\alpha$ using the
    isospin method alone, while the shaded region shows the
    result with the $SU(3)$ requirement as discussed in the text.} 
\label{fig:Alphaplot}
\end{center}
\end{figure}

We extract information on $\da \equiv\alphaeff - \alpha$  and $\alpha$
using isospin relations \cite{Isospin} that relate the decay amplitudes of 
$B\to\pi\pi$ decays and measurements of the branching fraction
and time-dependent \CP asymmetries in the decay $\Bz\to\pip\pim$ from
\babar\ \cite{pipiResults}.  For each of
the six observable quantities required to calculate $\alpha$
[$\BR(\Bz\to\pip\pim)$, \BR(\btopipiz), \BR(\Btopizpiz), 
${\cal  S}_{\pip\pim}$, ${\cal C}_{\pip\pim}$, and ${\cal
  C}_{\piz\piz}$], we generate an ensemble of simulated experiments
with uncorrelated Gaussian
distributions where the width on each distribution is the sum in
quadrature 
of the statistical and systematic errors of that measurement. Sets of
generated experiments that result in
an unphysical asymmetry or violate isospin are removed
from the sample.  Using the resulting distributions for \da and
$\alpha$, we calculate a confidence level (C.L.) for each solution and
plot the maximum value of $1$-C.L. of the various solutions in
Fig.~\ref{fig:Alphaplot}. One can further constrain $\alpha$ by using
the fact that the penguin amplitude contribution
to $\B\to\pi\pi$ decays must be very large if $\alpha$ is near 0 or $\pi$.
 We obtain a bound on the
magnitude of the penguin amplitude from the branching fraction of the
penguin-dominated decay 
$B_{s}\to\Kp\Km$ \cite{BstoKK} by making the conservative assumption
of $SU(3)$ breaking at less than $\sim100\%$ \cite{PengAmplitude}.
In Fig.~\ref{fig:Alphaplot} we also
show bounds on $\alpha$ when the size of the penguin amplitude is
constrained by this assumption.

In summary, we measure the branching fractions and \CP asymmetries in
\Btopizpiz, \Btopipiz, and \Btokpiz decays reconstructed from a sample
of approximately $383\times 10^{6}$ \BB pairs. All results are consistent
with previously published results~\cite{previous,previousBelle}, and
supersede the previous \babar\ results.  For
the $B\to\pi\pi$ 
decays, we find $\BR(\Btopizpiz) = (\Bzz \pm \Bzzstat \pm
\Bzzsyst)\times 10^{-6}$, $\BR(\Btopipiz) = (\Bppz \pm
\Bppzstat \pm \Bppzsyst)\times 10^{-6}$,  
$\Cpizpiz = \Czz \pm \Czzstat \pm \Czzsyst$, and 
$\acp_{\pipm\piz}   = \Appz\pm \Appzstat \pm \Appzsyst$.  We constrain
\da to be less than $39\degrees$ and exclude the range
$\left[25\degrees,66\degrees\right]$ in $\alpha$ at $90\%$ confidence level.  If
we consider only the preferred solution \cite{preffered}, we find 
$\alpha =(96^{+10}_{-6})\degrees$. For the \Btokpiz
decay, we find 
$\BR(\Btokpiz) 
= (\Bkpz \pm \Bkpzstat \pm \Bkpzsyst) \times 10^{-6}$ and
$\acp_{\Kpm\piz}   = \Akpz \pm \Akpzstat \pm \Akpzsyst$.  The
difference between $\acp_{\Kpm\piz}$ and  $\acp_{\Kpm\pimp} =
-0.107\pm 0.019$ \cite{pipiResults} indicates that
the effect of color-suppressed tree and electroweak penguin
amplitudes are significant.

We are grateful for the excellent luminosity and machine conditions
provided by our \pep2\ colleagues, 
and for the substantial dedicated effort from
the computing organizations that support \babar.
The collaborating institutions wish to thank 
SLAC for its support and kind hospitality. 
This work is supported by
DOE
and NSF (USA),
NSERC (Canada),
CEA and
CNRS-IN2P3
(France),
BMBF and DFG
(Germany),
INFN (Italy),
FOM (The Netherlands),
NFR (Norway),
MIST (Russia),
MEC (Spain), and
STFC (United Kingdom). 
Individuals have received support from the
Marie Curie EIF (European Union) and
the A.~P.~Sloan Foundation.


\begin{thebibliography}{99}
\bibitem{cpimplied} 
Unless specifically stated, conjugate decay modes are assumed
throughout this paper.
\bibitem{Argus} 
The function is $f(x)\propto x\sqrt{1-x^2}\exp{[-\zeta(1-x^2)]}$,
where $x=m_{\rm ES}/m_0$, $m_0$ is the \mes endpoint, and
$\zeta$ the shape parameter.
\end{thebibliography}
\end{document}